\documentclass{osa-article}
\usepackage{float}

\usepackage{wrapfig}
\usepackage{cite}
\usepackage{graphicx}
\usepackage{amsmath}
\usepackage{subfigure}
\journal{osajournal}

\setprjcopyright

\articletype{Research Article}

\begin{document}

\title{Powerful supercontinuum vortices generated by femtosecond vortex beams with thin plates}

\author{Litong Xu\authormark{1,$\dagger$}, Dongwei Li\authormark{2,$\dagger$}, Junwei Chang\authormark{2},  Deming Li\authormark{2}, Tingting Xi\authormark{1,3} and Zuoqiang Hao\authormark{2,4} }

\address{\authormark{1}School of Physical Sciences, University of Chinese Academy of Sciences, Beijing 100049, China\\

\authormark{2}Shandong Provincial Engineering and Technical Center of Light Manipulations \& Shandong Provincial Key Laboratory of Optics and Photonic Device, School of Physics and Electronics, Shandong Normal University, Jinan 250358, China\\
$^\dagger$ \emph{Authors contributed equally to this work.}
}

\email{\authormark{3} ttxi@ucas.ac.cn \authormark{4}zqhao@sdnu.edu.cn } 



\begin{abstract}
We demonstrate numerically and experimentally the generation of powerful supercontinuum vortices from femtosecond vortex beams by using multiple thin fused silica plates.
The supercontinuum vortices are shown to preserve the vortex phase profile of the initial beam for spectral components ranging from 500 nm to 1200 nm.
The transfer of the vortex phase profile results from the inhibition of multiple filamentation and the preservation of vortex ring with relatively uniform intensity distribution by means of the thin-plate scheme, where the  supercontinuum is mainly generated from the self-phase modulation and self-steepening effects.
Our scheme works for vortex beams with different topological charges, which provides a simple and effective method to generate supercontinuum vortices with high power.
\end{abstract}

\section{Introduction}
Vortex beams, characterized with ring shaped intensity distribution and helical wavefront with a phase singularity in the beam center \cite{sztul2006laguerre}, have attracted great attention since Allen \textit{et al} proposed their well-defined orbital angular momentum \cite{allen1992orbital}.
The orbital angular momentum that each photon of the vortex beam carries, is proportional to the topological charge and can be transferred between an object and the light field \cite{he1995direct}.
Based on the distinctive characteristics, vortex beams have found many important applications in various areas, such as optical tweezers \cite{grier2003revolution}, microscopy \cite{furhapter2005spiral}, and optical communications \cite{barreiro2008beating}.
On the other hand, the ultrashort pulse with spiral wavefront provides opportunities for new applications in high field physics, such as vortex THz generation \cite{Maksym2019Intensity}, electron vortices generation \cite{PhysRevA.96.043426}, proton acceleration \cite{2014Proton} and so on.
Femtosecond vortex beams can be generated from a femtosecond gaussian beam with phase encoding elements, such as spatial light modulator(SLM) \cite{neshev2010supercontinuum}, spiral phase plate \cite{Vuong2006} and q-plate \cite{marrucci2006optical}.
However, few-cycle or attosecond vortex beams are hard to realize by the above mentioned methods due to chromatic aberrations.
Supercontinuum vortices generated by nonlinear propagation of femtosecond vortex beams should be a good choice to generate few-cycle vortex beams. 

The supercontinuum vortices have been reported to be generated by the four-wave frequency mixing approach, proposed by Hansinger \textit{et al} \cite{Hansinger:14,hansinger2016white}.
However, this method cannot be used to generate high power supercontinuum vortices.
When intense femtosecond laser pulses propagate in Kerr medium,  supercontinuum emission are mainly generated by self-phase modulation(SPM) \cite{alfano1970observation} and self-steepening effects \cite{yang1984spectral}.
When filamentation occurs, ionization also plays an important role in the blue shift of spectrum broadening \cite{Zhan2018}. However, the supercontinuum will lose the vortex phase profile if multiple filaments are generated \cite{neshev2010supercontinuum,maleshkov2011filamentation}.
In this case, the multiple filaments which result from the initial perturbation are incoherent, and the phase of filaments accumulated from SPM effect destroys the regular distribution of the vortex phase profile. Consequently, the vortex phase profile cannot be transferred to the newly generated spectral components \cite{neshev2010supercontinuum,maleshkov2011filamentation}.
Only low power supercontinuum vortices are reported to be generated in air when the peak power of laser pulse is close to the critical power for self-focusing \cite{Zhang_2021}.
Up to date, it has not been reported that high power supercontinuum vortices can be generated by SPM during the nonlinear propagation of femtosecond vortex beams with peak power many times the critical power for self-focusing.

In this paper, for the first time to our best knowledge, we demonstrate the generation of supercontinuum vortices from intense femtosecond vortex beam by using several thin fused silica plates.
The thin-plate scheme has been successfully used to generate  supercontinuum from femtosecond Gaussian beams\cite{he2017high,lu2019greater,seo2020high}. Here we use the thin-plate setup to inhibit the breakup of the intense femtosecond vortex beams.
Before the multi-filamentation induced by intrinsic noise, the beam exits the plate. During the propagation in air, the vortex beam undergoes self-healing process of spatial distortions \cite{berge2009self} and then the beam enters the next plate. Consequently, a femtosecond vortex beam with super-broadened spectrum is obtained, even though the peak power of the femtosecond vortex beam is many times the critical power for self-focusing.

\section{Approach}

\subsection{Numerical simulation}
To numerically investigate the spatiotemporal evolution of femtosecond laser beam with a central wavelength of 800 nm in the thin-plate setup, we solve the generalized nonlinear Schr\"{o}dinger equation coupled with the electron density equation \cite{cheng2016supercontinuum}:

\begin{equation}
 \begin{aligned}
\frac{\partial{E}}{\partial z}=&\frac{i}{2 k_{0}} T^{-1} \nabla_{\perp} E+i\frac{\omega_{0}}{c} n_{2} T \int_{-\infty}^{t} \mathcal{R}\left(t-t^{\prime}\right)\left|E\left(t^{\prime}\right)\right|^{2} d t^{\prime} E\\
&+i \widehat{D} E-i \frac{k_{0}}{2 n_{0} \rho_{c}} T^{-1} \rho E-\frac{\beta^{(\kappa)}}{2}|E|^{2 \kappa-2} E-\frac{\sigma}{2} \rho E,
\end{aligned}
\label{E}
\end{equation}
\begin{equation}
 \frac{\partial \rho}{\partial t}=\frac{\beta^{(\kappa)}|E|^{2 \kappa}}{U_{i}}+\frac{\sigma \rho |E|^{2}}{U_{i}}-\frac{\rho}{\tau_{r e c}}.
 \label{ne}
\end{equation}

Here $E$ is the envelope of the electric field, and $z$ is the propagation distance. The operator $T=1+\frac{i}{\omega_{0}} \partial_{t}$ in the first term of the right side of Eq.\ref{E} accounts for space-time focusing.
$R(t)=(1-f) \delta(t)+f \Theta(t) \frac{1+\omega_{R}^{2} \tau_{R}^{2}}{\omega_{R} \tau_{R}^{2}} e^{-t / \tau_{R}} \sin \left(\omega_{R} t\right)$ denotes the Kerr nonlinear response, composed of an instantaneous part and a delayed Raman part, where the ratio $f$ is 0.18 for silica \cite{rolle2014filamentation} and 0.5 for air \cite{champeaux20083+}.
The operator $T$ before this term accounts for the self-steepening effect.
 $\widehat{D}=\sum_{n \geq 2}\left(\frac{k^{(n)}}{n !}\right)\left(i \partial_{t}\right)^{n}$ is the dispersion operator with $k^{(n)}=\partial^{n} k /\left.\partial \omega^{n}\right|_{\omega_{0}}$, where the dispersion relations of fused silica and air are taken from \cite{malitson1965interspecimen,zhang2008precision}. The critical plasma density is given by $\rho_{c} =1.72\times 10^{21} \ \mathrm{cm}^{-3}$. 
 The multiphoton absorption coefficient is taken to be $\beta_{\kappa} = 8.4\times 10^{-67}\rm\ cm^{9}/W^5$ for fused silica($\kappa=6$) and $\beta_{\kappa} = 3.1\times 10^{-98}\rm\ cm^{13}/W^7$  for air($\kappa=8$). Values of other parameters including the coefficients of nonlinear refractive index $n_2$ \cite{liu2005direct}, ionization potential $U_i$, avalanche cross section $\sigma$, and electron recombination time $\tau_{\text {rec }}$ \cite{xu2019supercontinuum} are listed in Table \ref{parameters}.

\begin{table}[h]
\begin{center}
\begin{tabular}{lcc}
\hline \hline \textbf{Parameters vs medium} & \textbf{Fused Silica} & \textbf{Air} \\
\hline
$n_{2}\left(10^{-19} \mathrm{~cm}^{2} / \mathrm{W}\right)$ & 3200 & 0.96 \\
$\tau_{R}(\mathrm{fs})$ & 32 & $70$ \\
$\omega_{R}^{-1}(\mathrm{fs})$ & $12.2$ & $62.5$ \\
$U_{i}(\mathrm{eV})$  & $7.8$ & $12.1$ \\
$\sigma\left(10^{-20} \mathrm{~cm}^{2}\right)$ & $65.7$ & $5.44$ \\
$\tau_{\text {rec }}(\mathrm{fs})$ & 150 & $+\infty$ \\
\hline \hline

\end{tabular}
\caption{Simulation parameters at 800 nm.}
\label{parameters}
\end{center}
\end{table}

The envelope of the vortex beam entering the first plate in cylindrical coordinate is given by
\begin{equation}
E(r,\varphi,t,z=0) = E_0(r/w)^{|m|}e^{-r^2/2w^2}e^{im\varphi}e^{-t^2/0.72\tau^2}e^{-ik_0r^2/2f}
\end{equation}

where the spatial distribution is a Lagarre-Gaussian mode with topological charge $m$, $w=200\ \mu$m, and 10\% random amplitude perturbation is introduced to the initial field \cite{xu2021helical}.
The pulse has a duration (FWHM) of $\tau=$ 50 fs, and energy of 420 $\mu$J.
The vortex beam is focused by a lens with the focal length $f$ = 11 cm.

The thicknesses of plates are varied to acquire stronger spectrum broadening while avoiding multiple filamentation. This is ensured by monitoring the transverse intensity distribution at peak of the pulse.
If the intensity fluctuation on the ring increases to about 20\%, the beam is released to the air.
Moreover, the peak intensity of the laser beam is kept lower than 20 TW/cm$^2$ to avoid the damage to fused silica \cite{cheng2016supercontinuum} . Because the coefficient of nonlinear refractive index in air is lower, the linear propagation of the vortex beam is dominant, and the intensity fluctuation will be decreased. Therefore, the spacings of plates are also adjusted by monitoring the transverse intensity distribution at peak of the pulse. The propagation in air will be terminated and the beam will enter next plate, when the intensity fluctuation on the ring decreases to about 10\% and the peak intensity of the laser beam is lower than 20 TW/cm$^2$.

\subsection{Experiment}
\begin{figure}[bp]
\centering

	\includegraphics[width=0.8\textwidth]
	{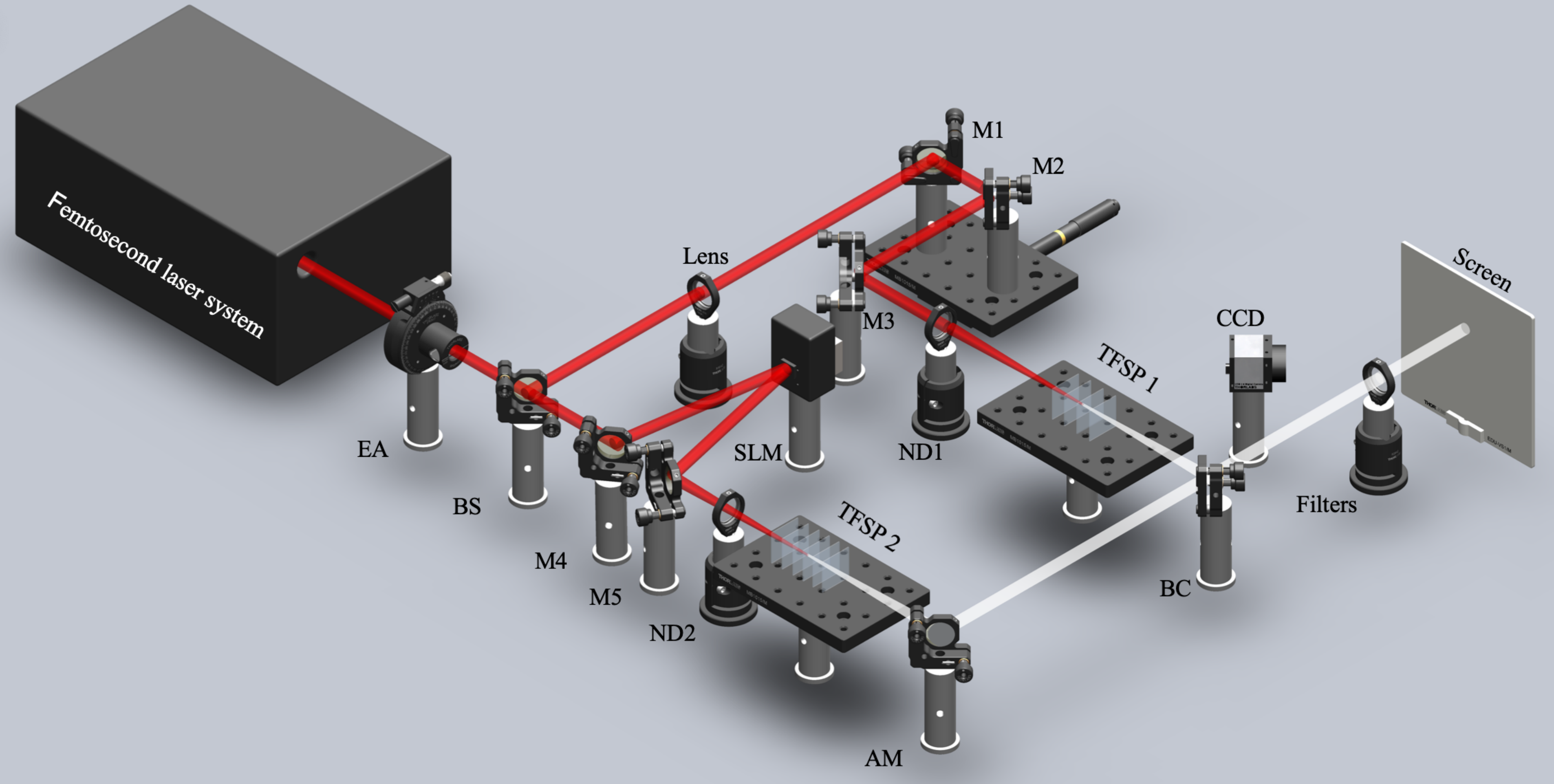}
\caption{Experimental setup. EA: Energy attenuator, BS: beam splitter, Lens: $f$=2 m, M1-M5:800 nm high reflective mirrors, SLM: spatial light modulator, ND: Neutral filter, TFSP: thin fused silica plates, AM: aluminum mirror, BC: beam combiner, Filters: suitable bandpass filters, short pass filters, and long-pass filters.}
\label{setup}
\end{figure}

The experimental setup is illustrated in Fig. \ref{setup}. 
The laser source is an amplified Ti: Sapphire femtosecond laser system (Solstice Ace, Spectra-physics) with a central wavelength of 800 nm, pulse duration of 50 fs, and repetition rate of 1 kHz.
Incident pulse energy is adjusted by the half-wave plate and ultra-broadband wire grid polarizers(WP25L-UB, Thorlabs Inc).
The linearly polarized Gaussian beam is reflected by a SLM to generate a vortex beam with different topological charges.
Note that a fresnel lens phase with focal length $f =2$ m is also encoded on the SLM.
The first fused silica plate is placed 11 cm before the focus for $m=1$ and 10 cm for $m=2$.
To avoid the nonlinear effect, the output supercontinuum is collected into an integrating sphere in the far field, and recorded by a spectrometer (USB- 4000, Ocean Optics Inc).
To identify the vortex phase, the supercontinuum beam enters a classical Mach Zehnder interferometer system, and interferes with the reference supercontinuum Gaussian beam generated by another thin-plate system.
The interference pattern is recorded by a CCD camera.
Several suitable bandpass filters, short pass filters, and long-pass filters are placed after the interference system to record the interference patterns in desired spectral regions.

\section{Results and discussion}

\subsection{Simulation}

\begin{figure}[htbp]
\centering

\includegraphics[width=\textwidth]{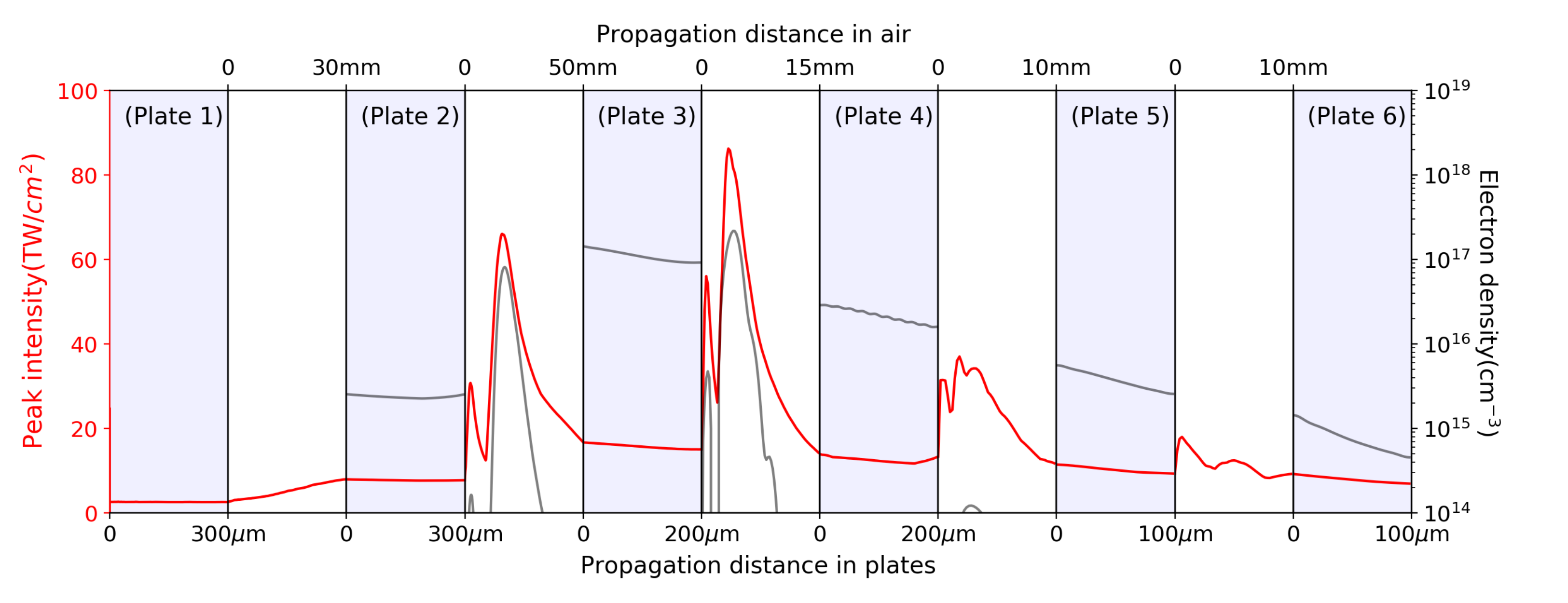}
\caption{Simulated evolution of peak intensity and electron density in the plates and air. The pulse energy of the femtosecond vortex beam is 420 $\mu$J, and topological charge $m=1$.}

\label{Imax}
\end{figure}

First, we simulate the propagation of a singly charged vortex beam in the thin-plate scheme.
Six thin fused silica plates are used. The thicknesses  of the plates are 300 $\mu$m, 300 $\mu$m, 200 $\mu$m, 200 $\mu$m, 100$\mu$m and 100 $\mu$m, respectively, and their spacings are 37 mm, 20 mm, 23 mm, 20 mm and 12 mm.
Figure \ref{Imax} shows the evolution of peak intensity of the vortex beam and electron density during the whole propagation in the plates and air.
It shows that the intensity in the six fused silica plates is kept lower than 20 TW/cm$^2$.
For the propagation in each fused silica, the peak intensity only changes slightly due to the short propagation distance.
The moderate magnitude of intensity avoids the damage of the fused silica and is high enough to trigger spectral broadening owing to the high nonlinear refractive index of fused silica. 
For the propagation in air, although the nonlinear effects are smaller, the vortex beam focuses firstly and then spreads out (see Fig. \ref{Imax}, the evolution in air between plate 2 and 3, plate 3 and 4).
As a consequence, the intensity in air increases firstly and then decreases.
This phenomenon has been observed for a Gaussian beam, and its mechanism is the focusing effect induced by nonlinear phase accumulated in the fused silica plates \cite{cheng2016supercontinuum,berge2009self}. 

\begin{figure}[htbp]
\centering

\includegraphics[width=\textwidth]
	{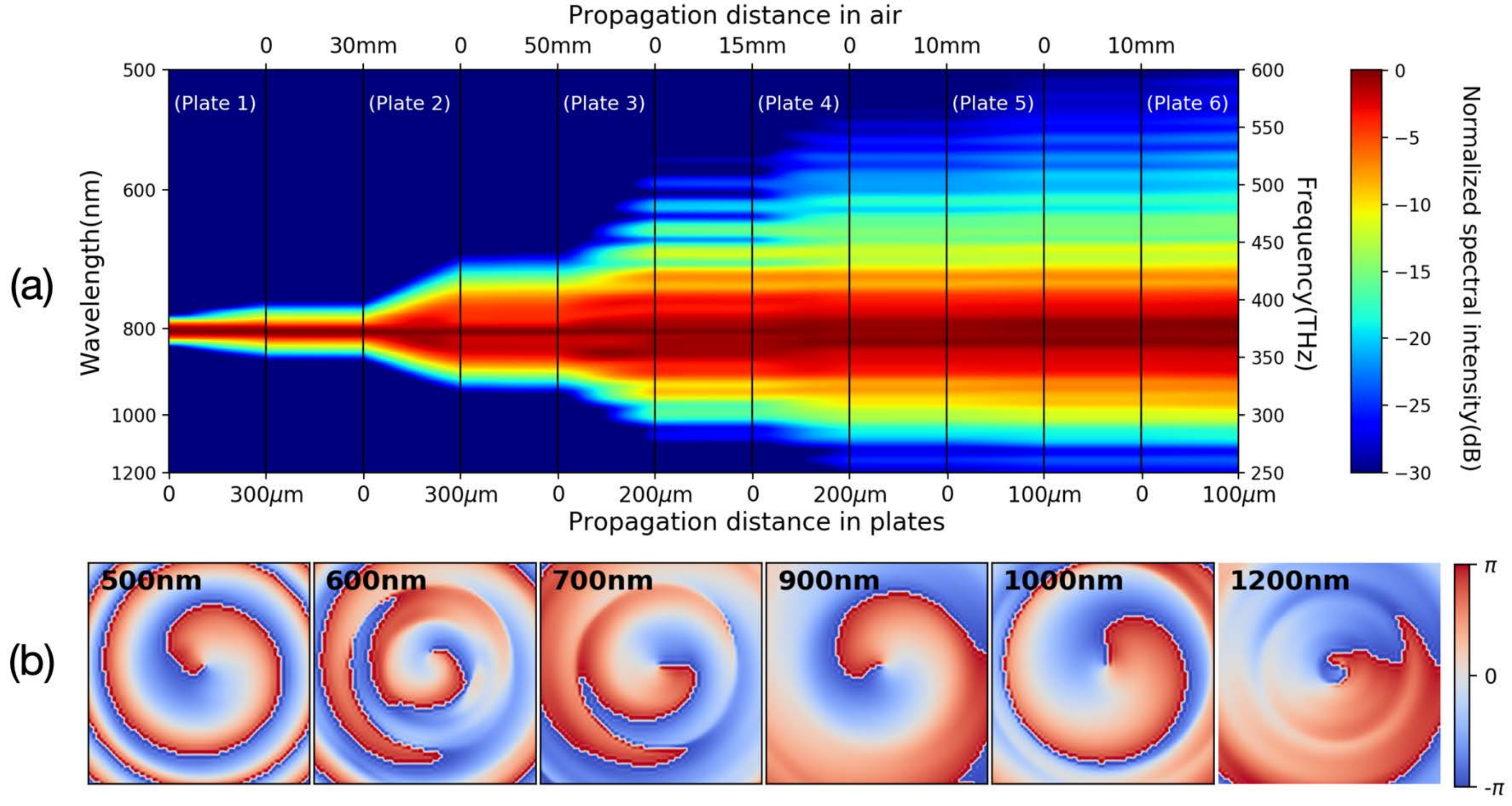}

\caption{(a) Simulated full-path spectrum evolution. (b) Transverse phase distribution of different spectral components for the femtosecond vortex beam with topological charge $m=1$.}
\label{spec-sim}
\end{figure}

The simulated spectra generated in the fused silica plates and air are shown in Fig. \ref{spec-sim}(a).
It can be seen that the spectrum is broadened only in the six fused silica plates, as we supposed.
During the propagation between two plates, the spectrum is hardly changed due to the less significant nonlinear effects in air.
In the first two plates, the obvious broadening occurs in both sides of red and blue shifts.
This symmetric broadening indicates that the main mechanism at this stage is SPM.
From the 3rd to the 6th plates, more blue shift is observed.
After the propagation of the 6th plate, the spectrum of the vortex beam covers from 500 nm to 1200 nm.
Adding more plates will not change the spectrum significantly, because the intensity drops quickly after the focal plane.
Figure \ref{spec-sim}(b) shows the simulated phase distribution of different spectral components of the generated supercontinuum in the transverse plane.
The spiral phase structure can be clearly seen in the transverse plane for wavelength from 500 nm to 1200 nm, where the phase variation along a closed path around the beam center is 2$\pi$.
It demonstrates that the supercontinuum generated by this thin-plate scheme preserves the phase singularity and the vortex phase profile of the initial beam.

When intense femtosecond laser pulses propagate in Kerr medium, supercontinuum is considered to be generated from SPM, electron generation and self-steepening effects.
To investigate the generation mechanism of the supercontinuum vortices, we should make clear which effects contribute to the spectral broadening significantly.
First, we compare the contributions of SPM and the electron generation.
The frequency shift induced by the two effects can be described by $\Delta \omega/\Delta z \propto \partial_t \rho(r, t)/2n_0\rho_c- n_2 \partial_t I(r, t)$.
From Fig. \ref{Imax}, we can see that during the propagation in the six plates, the electron density is relatively low.
We compare the frequency shift from each contribution in the 3rd plate, where the electron density is higher, and find that the frequency shift induced by electron generation is about two orders of magnitude smaller than that induced by SPM.
 \begin{figure}[htbp]
\centering

\includegraphics[width=0.8\textwidth]
	{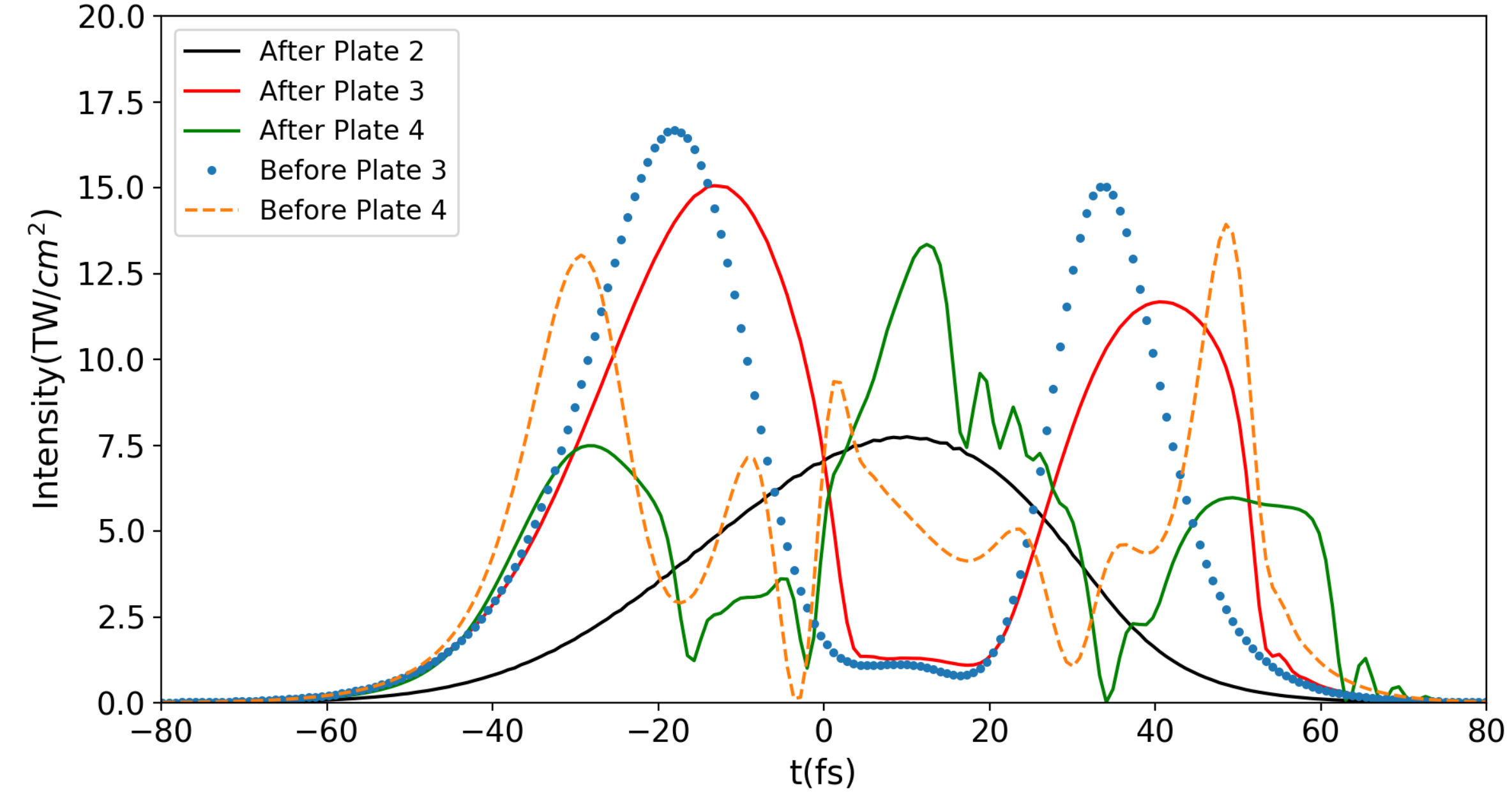}

\caption{The temporal profile of the femtosecond vortex beam at several longitudinal positions.}
\label{pulse}
\end{figure}
It indicates that the SPM contributes more significantly to the spectral broadening.
To investigate the influence of self-steepening effect, we plot the temporal profile of the femtosecond vortex beam at several longitudinal positions, as shown in Fig. \ref{pulse}.
For each longitudinal position, the transverse coordinate of the pulse is chosen by scanning the spatiotemporal intensity and finding its maximum.
As the pulse propagates in the first two plates, the self-steepening effect, which pushes the pulse peak to the tail, is not very distinct, thus the spectrum broadening mainly results from SPM, which leads to a symmetric spectrum broadening, as mentioned before.
When the femtosecond pulse propagates in the air between Plate 2 and Plate 3, the pulse splits into two sub-pulses without obvious steep trailing edge. After the 3rd plate, the steep trailing edge is observed in the pulse profile.
The self-steepening effect in the fused silica strengthens the blue extension of the spectrum by SPM, which is in accordance with the spectral evolution in the 3rd plate, as shown in Fig. \ref{spec-sim}.
These results suggest that both the SPM and self-steepening effects contribute to the supercontinuum generation.

\begin{figure}[htbp]
\centering

	\includegraphics[width=\textwidth]
	{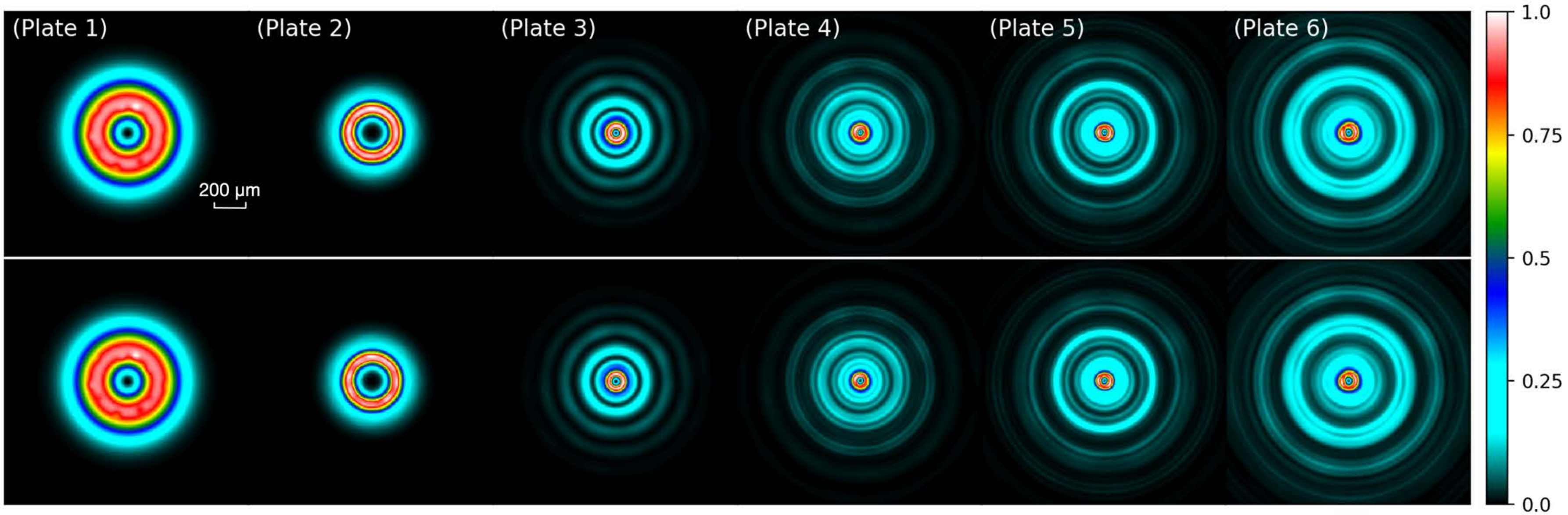}
\caption{Normalized transverse fluence distribution in the six plates. The first row denotes the front surface, and the second row denotes the back surface.}
\label{fluence}
\end{figure}
To explain the transfer of vortex phase from the initial vortex beam to the supercontinuum, we compare the energy fluence in the front and the back surface of the six plates, as shown in Fig. 5.
It shows that the fluctuation in the transverse plane will be enlarged after the propagation in each plate due to the modulation instability.
To be specific, the intensity fluctuation after the six plates increases to about 16\%, 13\%, 12\%, 18\%, 22\% and 17\%, respectively.
When propagating in air, since the peak power of the pulse is lower than the critical power for self-focusing, the linear effects dominate the propagation, and the intensity fluctuation in the vortex ring is decreased.
The intensity fluctuation decreases to 8\%, 5\%, 10\%, 13\% and 15\% before the 2nd to the 6th plates.
By use of the scheme, the vortex ring is kept with relatively uniform intensity distribution.
Based on the above discussion, the SPM and self-steepening effects contribute to the supercontinuum generation.
Both the two effects cause the phase accumulation of the new spectral components proportional to the intensity of the vortex beam.
As long as the intensity of the vortex ring is uniform, the newly spectral components, which are generated from the vortex ring, have the same phase distribution as the initial vortex beam.
Our scheme ensures the relatively uniform intensity distribution and enough spectral broadening.
Therefore, the vortex phase profile can be transferred to the supercontinuum without destructive interference.
Note that the nonlinear phase accumulated in previous plates \cite{berge2009self}, which induces the focusing of the beam, leads to the formation of concentric structure.
But this concentric structure will not prevent the transfer of vortex phase to the newly generated spectral components, because the resulted nonlinear phase accumulation for a certain radius is the same.
Consequently, supercontinuum vortices are generated, which have the same topological charge as the pump beam.

\subsection{Experiment}
\begin{figure}[htbp]
\centering

	\includegraphics[width=\textwidth]
	{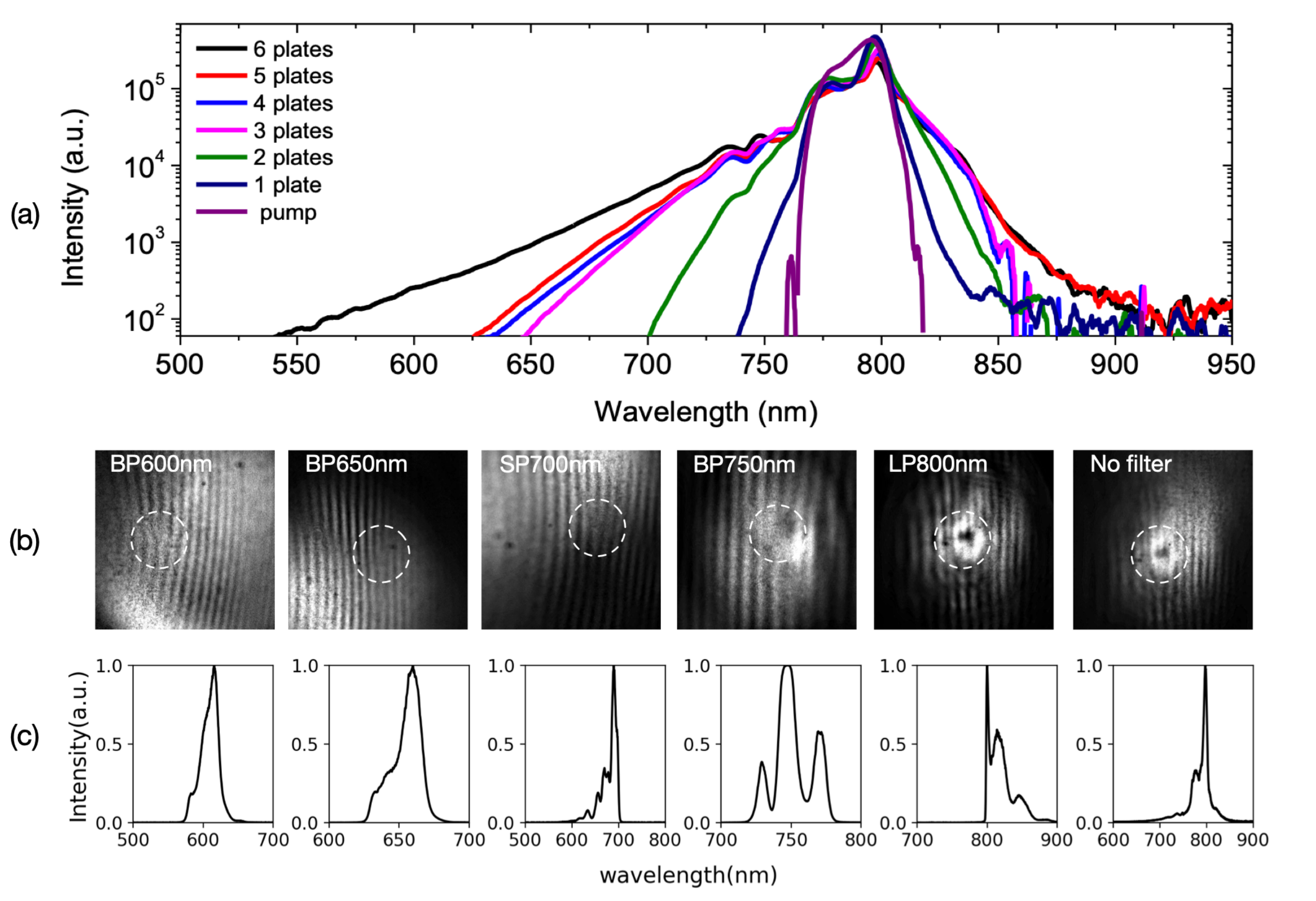}
\caption{(a) Experimentally measured spectrum after each plate. (b) Measured interference patterns after passing different filters. BP: band pass; SP: short pass; LP: long pass. (c) The detected spectra after different filters. The pulse energy before the first plate is 468 $\mu$J, the topological charge $m=1$.}
\label{exp-m1}
\end{figure}

To verify the scheme for generation of supercontinuum vortices, we experimentally investigate the propagation of singly charged vortex beam in the thin-plate scheme.
The number and thicknesses of the plates are the same as those in the simulation. 
The spacings of the plates are 37 mm, 20 mm, 23 mm, 20 mm and 12 mm, respectively.
The spectrum of the femtosecond vortex beam is broadened after each plate, as shown in Fig. \ref{exp-m1} (a).
The evolution of spectrum is similar to that in the simulation.
In the first two plates, the spectrum extends to both sides of red and blue shifts.
Starting from the 3rd plate, more blue extension is observed.
After propagating in the six plates, the spectrum extends to 550 nm in the blue extension. 
Figure \ref{exp-m1} (b) shows the interference patterns of different spectral components, which are obtained using suitable filters, and the spectra after corresponding filters are shown in Fig. \ref{exp-m1} (c).
The typical fork-like patterns are observed when using different filters, indicating that the vortex phase is transferred to the newly generated spectral components.
Moreover, the topological charge of the new spectral component is the same as the initial vortex beam.

\begin{figure}[htbp]
\centering

	\includegraphics[width=\textwidth]
	{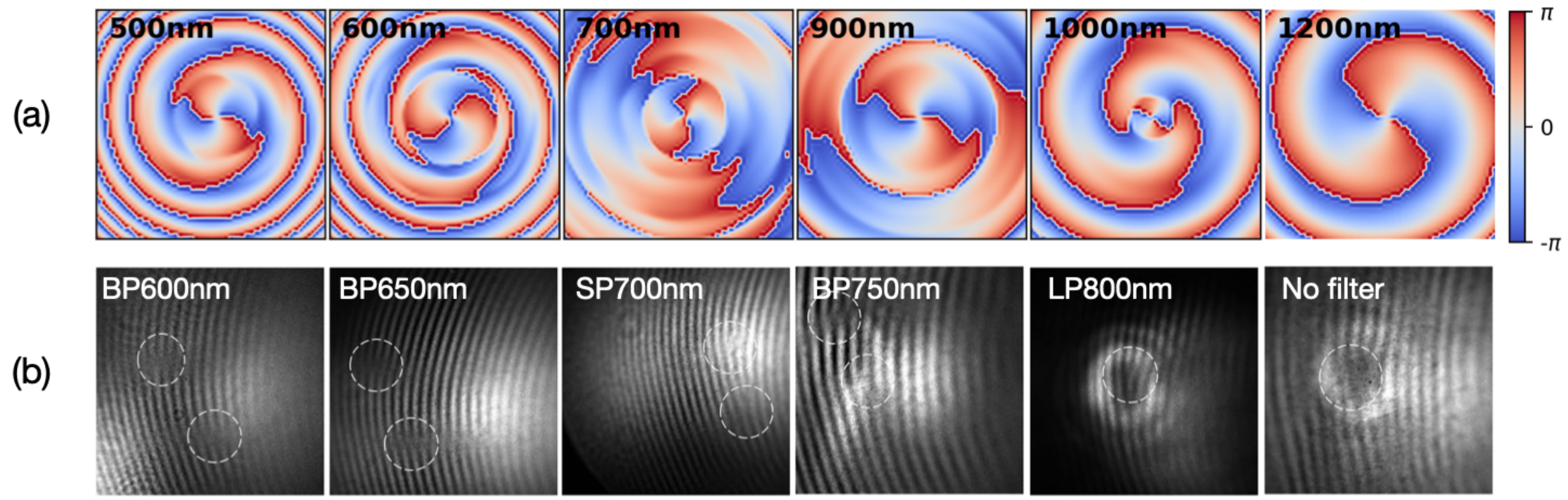}
\caption{ (a)Simulated transverse phase of different spectral components. (b) Measured interference patterns after passing different filters, the filters are the same with Fig. \ref{exp-m1}. The pulse energy before the first plate is 660 $\mu$J, the topological charge $m=2$. Five plates with thicknesses 300 $\mu$m, 300 $\mu$m, 200 $\mu$m, 200 $\mu$m and 100 $\mu$m are used. Their spacings are 18 mm ,14 mm,10 mm, and 16 mm.}
\label{exp-m2}
\end{figure}

Furthermore, we investigate the supercontinuum generation from the femtosecond vortex beam with double charge by using this thin-plate scheme.
Figure \ref{exp-m2} shows the simulated transverse phase and measured interference patterns of different spectral components.
We can see that the supercontinuum vortices with double topological charge can also be generated by this scheme, that is to say, our scheme works for vortex beams with different topological charges.
However, in this case, decay of high-order vortices is observed.
The intensity fluctuation induces phase distortion due to Kerr effect, leading to the break up of singularity, which is a common phenomenon when high nonlinearity is present \cite{hansinger2016white}.

\section{Concluison}

To summarize, we numerically and experimentally study the supercontinuum vortices generation from femtosecond vortex beams with high power by using the thin-plate scheme.
It is demonstrated that the ring shaped intensity distribution of the vortex beam can be kept after propagation in the thin plates, which enables the transfer of vortex phase distribution from fundamental frequency to newly generated ones for spectral components ranging from 500 nm to 1200 nm.
The SPM and self-steepening effect are responsible for the generation of the supercontinuum vortices in fused silica, and the propagation in air is responsible for the inhibition of beam collapse.
This scheme works for femtosecond vortex beams with different topological charges, providing a simple and effective method to obtain supercontinuum vortices with high power, which may be used to generate intense few-cycle vortex beams.

\section*{Funding}
National Natural Science Foundation of China (Nos. 11874056, 12074228, 11774038, 11474039), the Taishan  Scholar Project of Shandong Province (No.  tsqn201812043), and the Innovation Group of Jinan (Grant number: 2020GXRC039).

\section*{Disclosures}
The authors declare no conflicts of interest.

\bibliography{reference}

\begin{thebibliography}{10}
\newcommand{\enquote}[1]{``#1''}

\bibitem{sztul2006laguerre}
H.~Sztul, V.~Kartazayev, and R.~Alfano, \enquote{Laguerre-gaussian
  supercontinuum,} {\protect\JournalTitle{Optics Letters}} \textbf{31},
  2725--2727 (2006).

\bibitem{allen1992orbital}
L.~Allen, M.~W. Beijersbergen, R.~Spreeuw, and J.~Woerdman, \enquote{Orbital
  angular momentum of light and the transformation of laguerre-gaussian laser
  modes,} {\protect\JournalTitle{Physical Review A}} \textbf{45}, 8185 (1992).

\bibitem{he1995direct}
H.~He, M.~Friese, N.~Heckenberg, and H.~Rubinsztein-Dunlop, \enquote{Direct
  observation of transfer of angular momentum to absorptive particles from a
  laser beam with a phase singularity,} {\protect\JournalTitle{Physical Review
  Letters}} \textbf{75}, 826 (1995).

\bibitem{grier2003revolution}
D.~G. Grier, \enquote{A revolution in optical manipulation,}
  {\protect\JournalTitle{Nature}} \textbf{424}, 810--816 (2003).

\bibitem{furhapter2005spiral}
S.~F{\"u}rhapter, A.~Jesacher, S.~Bernet, and M.~Ritsch-Marte, \enquote{Spiral
  phase contrast imaging in microscopy,} {\protect\JournalTitle{Optics
  Express}} \textbf{13}, 689--694 (2005).

\bibitem{barreiro2008beating}
J.~T. Barreiro, T.-C. Wei, and P.~G. Kwiat, \enquote{Beating the channel
  capacity limit for linear photonic superdense coding,}
  {\protect\JournalTitle{Nature Physics}} \textbf{4}, 282--286 (2008).

\bibitem{Maksym2019Intensity}
Maksym, Ivanov, Illia, Thiele, Luc, Berg{\'e}, Stefan, Skupin, Danas, and
  Buoius, \enquote{Intensity modulated terahertz vortex wave generation in air
  plasma by two-color femtosecond laser pulses.} {\protect\JournalTitle{Optics
  letters}} \textbf{44}, 3889--3892 (2019).

\bibitem{PhysRevA.96.043426}
D.~Pengel, S.~Kerbstadt, L.~Englert, T.~Bayer, and M.~Wollenhaupt,
  \enquote{Control of three-dimensional electron vortices from femtosecond
  multiphoton ionization,} {\protect\JournalTitle{Phys. Rev. A}} \textbf{96},
  043426 (2017).

\bibitem{2014Proton}
X.~Zhang, B.~Shen, L.~Zhang, J.~Xu, X.~Wang, W.~Wang, L.~Yi, and Y.~Shi,
  \enquote{Proton acceleration in underdense plasma by ultraintense
  laguerre-gaussian laser pulse,} {\protect\JournalTitle{New Journal of
  Physics}} \textbf{16}, 123051 (2014).

\bibitem{neshev2010supercontinuum}
D.~N. Neshev, A.~Dreischuh, G.~Maleshkov, M.~Samoc, and Y.~S. Kivshar,
  \enquote{Supercontinuum generation with optical vortices,}
  {\protect\JournalTitle{Optics Express}} \textbf{18}, 18368--18373 (2010).

\bibitem{Vuong2006}
L.~T. Vuong, T.~D. Grow, A.~Ishaaya, A.~L. Gaeta, W.~Gert, E.~R. Eliel, and
  G.~Fibich, \enquote{Collapse of optical vortices,}
  {\protect\JournalTitle{Physical Review Letters}} \textbf{96}, 133901 (2006).

\bibitem{marrucci2006optical}
L.~Marrucci, C.~Manzo, and D.~Paparo, \enquote{Optical spin-to-orbital angular
  momentum conversion in inhomogeneous anisotropic media,}
  {\protect\JournalTitle{Physical review letters}} \textbf{96}, 163905 (2006).

\bibitem{Hansinger:14}
P.~Hansinger, G.~Maleshkov, I.~L. Garanovich, D.~V. Skryabin, D.~N. Neshev,
  A.~Dreischuh, and G.~G. Paulus, \enquote{Vortex algebra by multiply cascaded
  four-wave mixing of femtosecond optical beams,} {\protect\JournalTitle{Optics
  Express}} \textbf{22}, 11079--11089 (2014).

\bibitem{hansinger2016white}
P.~Hansinger, G.~Maleshkov, I.~Garanovich, D.~Skryabin, D.~Neshev,
  A.~Dreischuh, and G.~Paulus, \enquote{White light generated by femtosecond
  optical vortex beams,} {\protect\JournalTitle{Journal of the Optical Society
  of America B}} \textbf{33}, 681--690 (2016).

\bibitem{alfano1970observation}
R.~R. Alfano and S.~Shapiro, \enquote{Observation of self-phase modulation and
  small-scale filaments in crystals and glasses,}
  {\protect\JournalTitle{Physical Review Letters}} \textbf{24}, 592 (1970).

\bibitem{yang1984spectral}
G.~Yang and Y.~Shen, \enquote{Spectral broadening of ultrashort pulses in a
  nonlinear medium,} {\protect\JournalTitle{Optics Letters}} \textbf{9},
  510--512 (1984).

\bibitem{Zhan2018}
L.~Zhan, M.~Xu, T.~Xi, and Z.~Hao, \enquote{Contributions of leading and
  tailing pulse edges to filamentation and supercontinuum generation of
  femtosecond pulses in air,} {\protect\JournalTitle{Physics of Plasmas}}
  \textbf{25}, 103102 (2018).

\bibitem{maleshkov2011filamentation}
G.~Maleshkov, D.~N. Neshev, E.~Petrova, and A.~Dreischuh,
  \enquote{Filamentation and supercontinuum generation by singular beams in
  self-focusing nonlinear media,} {\protect\JournalTitle{Journal of Optics}}
  \textbf{13}, 064015 (2011).

\bibitem{Zhang_2021}
H.~Zhang, Y.~Zhang, S.~Lin, M.~Chang, M.~Yu, Y.~Wang, A.~Chen, Y.~Jiang, S.~Li,
  and M.~Jin, \enquote{Testing the coherence of supercontinuum generated by
  optical vortex beam in water,} {\protect\JournalTitle{Journal of Physics B:
  Atomic, Molecular and Optical Physics}} \textbf{54}, 165401 (2021).

\bibitem{he2017high}
P.~He, Y.~Liu, K.~Zhao, H.~Teng, X.~He, P.~Huang, H.~Huang, S.~Zhong, Y.~Jiang,
  S.~Fang \emph{et~al.}, \enquote{High-efficiency supercontinuum generation in
  solid thin plates at 0.1 tw level,} {\protect\JournalTitle{Optics Letters}}
  \textbf{42}, 474--477 (2017).

\bibitem{lu2019greater}
C.-H. Lu, W.-H. Wu, S.-H. Kuo, J.-Y. Guo, M.-C. Chen, S.-D. Yang, and A.~Kung,
  \enquote{Greater than 50 times compression of 1030 nm yb: Kgw laser pulses to
  single-cycle duration,} {\protect\JournalTitle{Optics Express}} \textbf{27},
  15638--15648 (2019).

\bibitem{seo2020high}
M.~Seo, K.~Tsendsuren, S.~Mitra, M.~Kling, and D.~Kim, \enquote{High-contrast,
  intense single-cycle pulses from an all thin-solid-plate setup,}
  {\protect\JournalTitle{Optics Letters}} \textbf{45}, 367--370 (2020).

\bibitem{berge2009self}
L.~Berg{\'e}, S.~Skupin, and G.~Steinmeyer, \enquote{Self-recompression of
  laser filaments exiting a gas cell,} {\protect\JournalTitle{Physical Review
  A}} \textbf{79}, 033838 (2009).

\bibitem{cheng2016supercontinuum}
Y.-C. Cheng, C.-H. Lu, Y.-Y. Lin, and A.~Kung, \enquote{Supercontinuum
  generation in a multi-plate medium,} {\protect\JournalTitle{Optics Express}}
  \textbf{24}, 7224--7231 (2016).

\bibitem{rolle2014filamentation}
J.~Rolle, L.~Berg{\'e}, G.~Duchateau, and S.~Skupin, \enquote{Filamentation of
  ultrashort laser pulses in silica glass and kdp crystals: A comparative
  study,} {\protect\JournalTitle{Physical Review A}} \textbf{90}, 023834
  (2014).

\bibitem{champeaux20083+}
S.~Champeaux, L.~Berg{\'e}, D.~Gordon, A.~Ting, J.~Pe{\~n}ano, and P.~Sprangle,
  \enquote{(3+ 1)-dimensional numerical simulations of femtosecond laser
  filaments in air: Toward a quantitative agreement with experiments,}
  {\protect\JournalTitle{Physical Review E}} \textbf{77}, 036406 (2008).

\bibitem{malitson1965interspecimen}
I.~H. Malitson, \enquote{Interspecimen comparison of the refractive index of
  fused silica,} {\protect\JournalTitle{Journal of the Optical Society of
  America}} \textbf{55}, 1205--1209 (1965).

\bibitem{zhang2008precision}
J.~Zhang, Z.~Lu, and L.~Wang, \enquote{Precision refractive index measurements
  of air, n2, o2, ar, and co2 with a frequency comb,}
  {\protect\JournalTitle{Applied Optics}} \textbf{47}, 3143--3151 (2008).

\bibitem{liu2005direct}
W.~Liu and S.~Chin, \enquote{Direct measurement of the critical power of
  femtosecond ti: sapphire laser pulse in air,} {\protect\JournalTitle{Optics
  Express}} \textbf{13}, 5750--5755 (2005).

\bibitem{xu2019supercontinuum}
M.~Xu, L.~Zhan, T.~Xi, and Z.~Hao, \enquote{Supercontinuum generation by
  femtosecond flat-top laser pulses in fused silica,}
  {\protect\JournalTitle{Journal of the Optical Society of America B}}
  \textbf{36}, G6--G12 (2019).

\bibitem{xu2021helical}
L.~Xu, D.~Li, J.~Chang, T.~Xi, and Z.~Hao, \enquote{Helical filaments array
  generated by femtosecond vortex beams with lens array in air,}
  {\protect\JournalTitle{Results in Physics}} \textbf{26}, 104334 (2021).

\end{thebibliography}

\end{document}